\documentclass[aps,showpacs,amssymb,amsfonts,superscriptaddress,twocolumn,pra,nofootinbib]{revtex4-1}

\usepackage{bbm,mathrsfs}
\usepackage{graphicx}
\usepackage{amsfonts,amsmath}
\usepackage{amsthm}

\def\textbf#1{{\bf #1}}
\def\be{\begin{equation}}
\def\ee{\end{equation}}
\def\ben{\begin{eqnarray}}
\def\een{\end{eqnarray}}
\def\eea{\end{array}}
\def\bea{\begin{array}}
\newcommand{\Tr}[0]{\mathrm{Tr}}
\newcommand{\ot}[0]{\otimes}
\newcommand{\bei}{\begin{itemize}}
\newcommand{\eei}{\end{itemize}}
\newcommand{\ket}[1]{|#1\rangle}
\newcommand{\bra}[1]{\langle#1|}

\newcommand{\proj}[1]{\ket{#1}\!\bra{#1}}

\newcommand{\tri}[0]{\widetilde{r}}

\newcommand{\Eins}{\ensuremath{\mathbbm 1}}

\usepackage{changes}

\definecolor{myurlcolor}{rgb}{0,0,0.7}
\definecolor{myrefcolor}{rgb}{0.8,0,0}
\usepackage{hyperref}
\hypersetup{colorlinks, linkcolor=myrefcolor,
citecolor=myurlcolor, urlcolor=myurlcolor}

\newtheorem{thm}{Theorem}
\newtheorem{thmm}{Theorem}

\theoremstyle{definition}

\begin{document}

\title{Four-qubit entangled symmetric states with positive partial transpositions}

\author{J. Tura}

\affiliation{ICFO--Institut de Ci\`{e}ncies Fot\`{o}niques, 08860
Castelldefels (Barcelona), Spain}

\author{R. Augusiak}

\affiliation{ICFO--Institut de Ci\`{e}ncies Fot\`{o}niques, 08860
Castelldefels (Barcelona), Spain}

\author{P. Hyllus}
\affiliation{Department of Theoretical Physics, The University of
the Basque Country, P.O. Box 644, E-48080 Bilbao, Spain}

\author{M. Ku\'s}
\affiliation{Center for Theoretical Physics, Polish Academy of
Sciences, Al. Lotnik\'ow 32/46, 02-668 Warszawa, Poland}

\author{J. Samsonowicz}
\affiliation{Faculty of Mathematics and Information Science,
Warsaw University of Technology, Pl. Politechniki 1, 00-61
Warszawa, Poland}

\author{M. Lewenstein}

\affiliation{ICFO--Institut de Ci\`{e}ncies Fot\`{o}niques, 08860
Castelldefels (Barcelona), Spain} \affiliation{ICREA--Instituci\'o
Catalana de Recerca i Estudis Avan\c{c}ats, Lluis Companys 23,
08010 Barcelona, Spain}

\begin{abstract}
We solve the open question of the existence of four-qubit
entangled symmetric states with positive partial transpositions
(PPT states). We reach this goal with two different approaches.
First, we propose a half-analytical-half-numerical method that
allows to construct multipartite PPT entangled symmetric states
(PPTESS) from the qubit-qudit PPT entangled states. Second, we
adapt the algorithm allowing to search for extremal elements in
the convex set of bipartite PPT states [J. M. Leinaas, J. Myrheim,
and E. Ovrum,
\href{http://pra.aps.org/abstract/PRA/v76/i3/e034304}{Phys. Rev. A
{\bf 76}, 034304 (2007)}] to the multipartite scenario. With its
aid we search for extremal four-qubit PPTESS and show that
generically they have ranks $(5,7,8)$. Finally, we provide an
exhaustive characterization of these states with respect to their
separability properties.
\end{abstract}

\keywords{} \pacs{}

\maketitle

{\it Introduction.--} Entanglement has become an important notion
in modern physics \cite{HReview}. This striking feature of
composite physical systems not only fundamentally distinguishes
classical and quantum theories, but it has also developed into a
key resource for various applications. For instance, it allows for
quantum teleportation \cite{teleportation}, quantum cryptography
\cite{Ekert}, and is a prerequisite for another important resource
in quantum information theory (QIT)--nonlocal correlations
\cite{Barrett}. Deciding, then, if a given quantum state is
entangled (i.e., if it is not a mixture of products of states
representing individual subsystems \cite{separable}) has become
one of the most important problems (the so-called separability
problem) in QIT and, even if simple to formulate, it is one of the
hardest to solve \cite{Gurvits}.

Due to the recent achievements in experimental implementations of
various many-body states such as, for instance, the four-qubit
bound entangled Smolin state \cite{SmolinExp}, the six-qubit Dicke
state states \cite{DickeExp}, or the eight-qubit Greenberger-
Horne-Zeilinger (GHZ) state \cite{GHZExp}, the separability
problem in quantum systems consisting of more than two
constituents has gained importance. Here, the problem becomes even
more complicated because one wants to answer not only the simple
question of whether a particular state is entangled, but also what
type of entanglement it has (see Ref. \cite{DurCirac}). Various
approaches have been proposed to detect and characterize
entanglement in such systems (see, e.g., Refs.
\cite{multWit,DIW,SpinSq,SeevinckUffinkGuhne} and a recent review
\cite{GezaReview}).

With this paper we fit into the above line of research and start a
general program of characterization of entanglement properties and
correlations of an important class of multipartite states---the
so-called symmetric states\footnote{In other words, states
describing bosonic systems consisting of finite number of
two-dimensional subsystems. Here we assume the usual definition of
separability, with respect to the Hilbert space being a product of
single-particle Hilbert spaces. Other definitions are also
considered in such systems (see, e.g., Ref. \cite{Benatti} and
references therein).}. These states have already been investigated
(see, e.g., Refs. \cite{EckertAnP,sym,Perm1,Perm2,symm}). More
attention, however, has been devoted to pure states, while
entanglement properties of mixed states are mostly unstudied. In
particular, it remains uncertain if there exist four-qubit
entangled symmetric states with all partial transpositions
positive or, in other words, whether the separability condition
based on partial transposition is necessary and sufficient in this
case. It is known that PPT symmetric states of three-qubits are
all separable \cite{EckertAnP}, and existence of such states of
five and six qubits has recently been reported \cite{Perm1,Perm2}.
The main aim of the paper is to fill in this gap by showing,
contrary to common belief, that there exist four-qubit PPTESSs.
Then, we thoroughly study the entanglement properties of
four-qubit PPT symmetric states.

%

{\it Preliminaries and general entanglement properties of
four-qubit symmetric states.--} Let us start from a couple of
definitions that will frequently be used throughout the paper.
Consider a product Hilbert space $H_{d,N}=(\mathbbm{C}^d)^{\ot
N}$ and a convex set $D$ of $N$-partite states $\rho$ acting
on $H_{d,N}$. By $r(\rho)$, $K(\rho)$, and $R(\rho)$ we
will be denoting the rank, kernel, and range of $\rho$. Also,
the notations $(f_0,\ldots,f_{d-1})$ and $\ket{e_1,\ldots,e_N}$
will be used to denote a vector
$\ket{f}\in\mathbbm{C}^d$ and a pure product vector from
$H_{d,N}$, respectively.

We say that $\rho$ is {\it fully separable} iff it can be written
in the following form \cite{separable}:
\begin{equation}\label{separable}
\rho=\sum_{i}p_i\rho_{A_1}^{i}\ot\ldots\ot\rho_{A_{N}}^{i},\quad p_i\geq0,\quad \sum_ip_i=1,
\end{equation}
where $A_1,\ldots,A_N$ denote parties and $\rho_{A_{j}}^{i}$
are density matrices representing respective subsystems.

Then, splitting the parties $A_1,\ldots,A_N$ into two disjoint
subsets $S$ and $\overline{S}$, we say that $\rho$ is PPT with
respect to the bipartition $S|\overline{S}$ if and only if
$\rho^{T_{S}}\geq 0$ (note that the partial transpositions with
respect to $\overline{S}$ and $S$ are equivalent under the full
transposition). States with this property make a convex subset
$D_S$ of $D$. An intersection of these subsets for all $S$
together with $D$ defines the set, denoted $D_{\mathrm{PPT}}$, of
states that have all partial transposes positive, further called
{\it PPT states}. An important class of PPT entangled states are
the so-called edge states \cite{edge,Karnas}. We say that a given
PPT state $\rho$ is {\it edge} if there is no product vector
$\ket{e_1,\ldots,e_N}\in H_{d,N}$ such that
$\ket{e_1,\ldots,e_N}\in R(\rho)$ and its partial conjugation with
respect to $S$ is in $R(\rho^{T_S})$ for all $S$. Edge states are
an important tool in the characterization of PPT entangled states
because every PPT state can be decomposed into a convex
combination of an edge state and a separable state, or, more
precisely, edges states are those from which no separable states
can be subtracted without loosing the PPT property. Thus, they are
crucial for the full characterization of entanglement in PPT
states. More attention has been devoted to edge states in biparite
and three-partite systems (see Refs. \cite{edgePapers,Karnas}),
while little is known about them in $N$-partite systems. A very
convenient way to classify and characterize edge states is to use
their ranks together with the rank of all the relevant partial
transposes, i.e.,
$(r(\rho),r(\rho^{T_{A_1}}),\ldots,r(\rho^{T_{A_N}}),r(\rho^{T_{A_1A_2}}),\ldots)$.

We can now pass to the $N$-qubit symmetric states. Let us focus on
$H_{2,N}$ and denote by $S_N$ and $P_N$ the symmetric subspace of
$H_{2,N}$ and a projector onto $S_N$. Recall, that $S_N$ is
spanned by the unnormalized vectors
$\ket{E_i^{N}}=\ket{\{0,N-i\},\{1,i\}}$, where
$\ket{\{0,N-i\},\{1,i\}}$ is a symmetric vector consisting of $i$
ones and $N-i$ zeros. For further benefits, let us notice that
$\mathrm{dim}S_N=N+1$ and hence $S_N$ is isomorphic to
$\mathbbm{C}^{N+1}$. We then call a density matrix $\rho$ acting
on $H_{2,N}$ {\it symmetric} iff $R(\rho)\subseteq S_N$.

In the case of symmetric states the number of relevant partial
transpositions defining the set of PPT states can be significantly
reduced. Clearly, positivity of a partial transposition with
respect to a particular $S$ implies positivity of partial
transposition with respect to all subsets $S$ with the same number
of parties. Together with the equivalence of some of partial
transpositions under the full transposition, this results in only
$\lfloor N/2\rfloor$ of relevant partial transpositions. For
concreteness, we choose them to be $T_{A_1}$, $T_{A_1A_2}$,
$T_{A_1A_2A_3}$, etc. In the particular case of $N=4$ there are
only two of them, which, breaking the general notation, we will be
denoting by $T_A$ and $T_{AB}$. Consequently, the set of
four-qubit PPT symmetric states $D_{\mathrm{PPT}}^{\mathrm{sym}}$
is an intersection of three sets $D^{\mathrm{sym}}$,
$D^{\mathrm{sym}}_{A}$, and $D^{\mathrm{sym}}_{AB}$. Accordingly,
one has in this case only three relevant ranks
$(r(\rho),r(\rho^{T_A}),r(\rho^{T_{AB}}))$, which, for the sake of
simplicity, we call {\it three-rank} of $\rho$ and denote as
$\tri(\rho)$.  Notice also that the fact that $\rho$ is symmetric
imposes nontrivial bounds on $r(\rho)$, $r(\rho^{T_A})$, and
$r(\rho^{T_{AB}})$. First of all, $\mathrm{dim}S_4=5$ implies
$r(\rho)\leq 5$. Then, since $S_3$ is isomorphic to
$\mathbbm{C}^4$, while $S_2$ to $\mathbbm{C}^3$,
$r(\rho^{T_A})\leq 8$ and $r(\rho^{T_{AB}})\leq 9$.

Passing to the separability properties of PPT qubit symmetric
states, it is known that for $N=2$ and $N=3$ all are separable.
While the first case directly follows from the results of
\cite{Horodecki96PLA}, for the second one, one uses the fact that
$S_2$ is isomorphic to $\mathbbm{C}^3$ and therefore $\rho$ can be
seen as a PPT qubit-qutrit state. Again, the results of Ref.
\cite{Horodecki96PLA} apply here. Finally, it was shown in Ref.
\cite{EckertAnP} that if $\rho$ is a symmetric $N$-qubit state and
$r(\rho)\leq N$, then it takes the form (\ref{separable}). The
first nontrivial, and so far unsolved case appears for $N=4$. All
PPT four-qubit symmetric states with $r(\rho)\leq 4$ are
separable, however, it has not been known whether the same holds
for $r(\rho)=5$. In other words, it remains uncertain whether
there are no PPT entangled four-qubit symmetric states and the
partial transposition provides a necessary and sufficient
criterion in this case. Our main aim is to show that this is not
the case and there do exist examples of PPT entangled states
supported on $S_4$.

Before getting to the construction, let us first discuss
separability properties of four-qubit PPT symmetric states and
single out all instances with respect to the three-rank when there
are edge states. All the theorems proven below are left with
sketches of proofs, while their detailed versions may be found in
appendix \ref{appA} and Ref. \cite{Jordi}.

Together with the already mentioned results of Ref.
\cite{EckertAnP} we have the following theorem.

\begin{thm}\label{thm_sep}
Let $\rho$ be a four-qubit PPT symmetric state. If either
$r(\rho)\leq 4$, or $r(\rho^{T_A})\leq 4$, or
$r(\rho^{T_{AB}})\leq 3$ then $\rho$ is separable, while if
$r(\rho^{T_A})\leq 6$, or $r(\rho^{T_{AB}})\leq 6$, then
generic $\rho$ of such ranks is separable.
\end{thm}

\begin{proof}The cases of $r(\rho)\leq 4$ and of $r(\rho^{T_A})\leq 4$ are
proven in Ref. \cite{EckertAnP}. The remaining ones follow from
the results of Refs. \cite{2N,MN}, which say that a PPT state
$\rho$ acting and supported on $\mathbbm{C}^{M}\ot\mathbbm{C}^{N}$
$(M\leq N)$ of rank $r(\rho)\leq N$ is separable. In the case of
$r(\rho^{T_{AB}})\leq 3$ one treats $\rho^{T_{AB}}$ as a PPT state
acting on $(\mathbbm{C}^{3})^{\ot 2}$. In the cases of
$r(\rho^{T_A})\leq 6$ and of $r(\rho^{T_{AB}})\leq 6$ one sees
$\rho^{T_A}$ and $\rho^{T_{AB}}$ as bipartite PPT states acting
and generically supported on $\mathbbm{C}^2\ot\mathbbm{C}^6$.
\end{proof}

\begin{thm}\label{thm_edge}
Generic four-qubit PPT symmetric states of a given three-rank
different from $(5,7,7)$ and $(5,7,8)$ are not edge.
\end{thm}
\begin{proof}
Roughly speaking, for a generic symmetric PPT state $\rho$ of a
particular three-rank $\widetilde{r}(\rho)$, except for $(5,7,7)$
and $(5,7,8)$, we find a symmetric product vector $\ket{e}^{\ot
4}\in H_{2,4}$ in the support of $\rho$ such that
$\ket{e^{*}}\ket{e}^{\ot 3}\in R(\rho^{T_A})$, and
$\ket{e^{*},e^{*},e,e}\in R(\rho^{T_{AB}})$. Clearly, due to
theorem \ref{thm_sep} most of the cases with respect to
$\widetilde{r}(\rho)$ are already ruled out, $(5,8,9)$ is trivial,
and those that need to be treated separately are $(5,8,7)$,
$(5,8,8)$, $(5,7,9)$ (see appendix \ref{appA}).
\end{proof}

Although using the above method we cannot prove that generic
states of ranks (5,7,7) are not edge, it is conjectured to be the
case. More importantly, there is an indication that they are
generically separable, but this will be studied elsewhere
\cite{Jordi}. Also, exploiting the methods below we obtained
examples of edge PPTESSs of ranks (5,7,8); all the found examples
of ranks (5,7,7) were separable.

Finally, let us prove that any entangled element of
$D^{\mathrm{sym}}_{\mathrm{PPT}}$ can be decomposed in terms of at
most six vectors of Schmidt rank two, i.e., entangled vectors that
can be written as a sum of two fully product vectors.

\begin{thm}\label{thm_dec}
Any entangled four-qubit symmetric state $\rho$ can be written as
\begin{equation}\label{Lemma1}
\rho=\sum_{k=1}^{K} [A_k (a_k,b_k)^{\ot 4}+B_k (a_k,-b_k)^{\ot
4}],
\end{equation}
where $K\leq 6$, $(a_k,b_k)\in\mathbbm{C}^{2}$,
$A_k,B_k\in\mathbbm{C}$, and by $[\psi]$ we denote a projector
onto $\ket{\psi}$.
\end{thm}
\begin{proof}
Applying a nonsingular transformation $V(a,b)^{\ot 2}=(a^2,b^2)$
with $a,b\in\mathbbm{C}$ to the last two qubits of a four-qubit
symmetric PPT entangled state $\rho$, one brings it to a
three-qubit PPT state $\sigma$ acting on $S_2\ot \mathbbm{C}^2$.
The latter is clearly separable and therefore can be written as a
convex combination of at most six product rank-one projections
\cite{Normal}, i.e.,
\begin{equation}
\sigma=\sum_{k=1}^{K}\proj{e_k}\ot\proj{f_k},
\end{equation}
where the form of $\ket{e_k}\in S_2$ can determined from the
orthogonality to $K(\sigma)$ and is given by
$\ket{e_k}=A_k(a^2_k,a_kb_k,b_k^2)+B_k(a^2_k,-a_kb_k,b_k^2)$,
while $\mathbbm{C}^2\ni\ket{f_k}=\ket{a^2_k,b^2_k}$ with
$a_k,b_k,A_k,B_k\in\mathbbm{C}$. To obtain (\ref{Lemma1}) and
complete the proof, one applies another full rank
transformation 
$W(a,b)\ot (a^2,b^2)=(a,b)^{\ot 3}$ $a,b\in\mathbbm{C}$
to the last two qubits of $\sigma$.
\end{proof}

Using the methods developed in Ref. \cite{Normal}, one can obtain
similar decomposition in which vectors $(a_k,-b_k)$ are replaced
by $(0,1)$ or $(1,0)$ (see appendix \ref{appA} for the proof).

%

{\it Constructing four-qubit PPT entangled symmetric states.--} We
start by introducing a class of qubit-qudit PPT entangled states
being a direct generalization of the $2\otimes 4$ PPT entangled
states introduced by Horodecki \cite{Horodecki97PLA} (see Ref.
\cite{Chruscinski} for generalizations of $3\ot 3$ Horodecki
states). To this end, consider the density matrices
\begin{equation}\label{states0}
\rho_\text{insep}^{d}=\frac{2}{2d-1}\sum_{i=0}^{d-2}
\proj{\Psi_i}+\frac{1}{2d-1}\proj{1,0},
\end{equation}
where $\ket{\Psi_i}=(1/\sqrt{2})(\ket{0,i}+\ket{1,i+1})$
$(i=0,\ldots,d-2)$. Then, analogously to \cite{Horodecki97PLA},
for any $d\geq 2$, we define
\begin{equation}\label{states}
\rho_{d,b}=\frac{[(2d-1)b\rho^d_{\mathrm{insep}}+\proj{\Phi_b}]}{(2d-1)b+1}\qquad
(0\leq b\leq 1)
\end{equation}
with
$\ket{\Phi_b}=(1/\sqrt{2})\ket{0}(\sqrt{1-b}\ket{0}+\sqrt{1+b}\ket{d-1})$.
For $d=4$, Eq. (\ref{states}) gives the original states of Ref.
\cite{Horodecki97PLA}.

Let us briefly characterize this class, and show that it maintains
the separability and PPTness properties of the original $2\otimes
4$ Horodecki state (see appendix \ref{appB} for more details).
First, one checks that for $d\geq 2$ and $b\in[0,1]$,
$\rho_{d,b}^{T_A}=(\mathbbm{1}_2\ot U)\rho_{d,b}(\mathbbm{1}_2\ot
U^{\dagger})$, where $U$ is an antidiagonal unitary operation
consisting of unities, and therefore $\rho_{d,b}$ is PPT. Second,
following the argumentation of Ref. \cite{Horodecki97PLA}, one can
show (cf. appendix \ref{appB}) 
that for $d\geq 4$ and $b\in(0,1)$, the states (\ref{states}) are
PPT entangled, and more importantly, they are edge, while for
$b=0$ or $b=1$, or $d=2,3$ separable. Third, one has that
$r(\rho_{d,b})=r(\rho_{d,b}^{T_{A}})=d+1$.

With the aid of the states $\rho_{d,b}$ we can construct PPT
symmetric entangled states. We will do that in few steps. First,
we apply a full--rank transformation
$F=\mathbbm{1}_d-y\ket{0}\!\bra{d-1}$ with $y=\sqrt{(1-b)/(1+b)}$
to the second subsystem of $\rho_{d,b}$ so that the product
vectors in the range of the resulting (unnormalized) state
$\rho'_{d,b}=(\mathbbm{1}\ot F)\rho_{d,b}(\mathbbm{1}\ot
F^{\dagger})$ are given by
\begin{eqnarray}\label{product}
(1,\alpha)&\ot& (\alpha^{d-1},\ldots,\alpha,1)\qquad (\alpha\in\mathbbm{C})\\
(0,1)&\ot&(1,0,\ldots,0).
\end{eqnarray}
The above form of the product vectors in $R(\rho'_{d,b})$ is a key
feature here because it allows one for a simple mapping of our
states to many-qubit symmetric states.

Second, to the same subsystem, we apply a nonsingular $d'\times d$ $(d'<d)$ matrix
\begin{equation}\label{N}
F_2=\sum_{i=0}^{d'}\sum_{j=0}^{d-d'}\gamma_j\ket{i}\!\bra{i+j},
\end{equation}
where $\gamma_i$ $(i=0,\ldots,d-d')$ are some complex parameters.
By doing so, we obtain another class of PPT states
$\rho_{d',b}''=(\mathbbm{1}\ot F_2)\rho'_{d,b}(\mathbbm{1}\ot
F_2^{\dagger})$ which act on $\mathbbm{C}^2\ot\mathbbm{C}^{d'}$
with $d<d'$ but have additional $d-d'+1$ parameters $\gamma_i$.
The transformation $F_2$ is chosen in such a way that it allows to
introduce additional parameters preserving the form of the product
vectors in the range of the resulting states. Precisely, the
product vectors in $R(\rho_{d',b}'')$, although living in a
smaller-dimensional Hilbert space, are of the form
(\ref{product}). Noticeably, since we are only using local
operations, the resulting states are also edge.

Third, we apply yet another local operation, denoted $V$, which
maps the local vectors $(\alpha^{d-1},\ldots,\alpha,1)$ to
$(1,\alpha)^{\ot (d-1)}$. Clearly, $V$ is of full rank because in
both vectors the same powers of $\alpha$ appear. By applying $V$
to the second subsystem of $\rho_{d',b}''$ we simply obtain
$(N=d')$-qubit symmetric states $\omega_{N}$, which by the very
construction, have all one-particle partial transpositions
positive.

As a result the above three filters allow us to construct a family
of many-qubit symmetric entangled states from the generalized
Horodecki states, which are our starting point for searching for
PPT entangled symmetric states. However, the states $\omega_{N}$
have in general nonpositive partial transpositions except for the
single-partite ones. To overcome this, we can consider another
state $\omega_{N,\lambda}=\omega_{N}+\lambda P_N$, where $P_N$
stands for the projector onto the symmetric subspace $S_N$.
Clearly there exists the smallest $\lambda$, denoted $\lambda_*$,
such that $\omega_{N,\lambda_{*}}$ is PPT. Although it keeps the
rank of the state constant, this operation, however, inevitably
increases the ranks of 
the partial transpositions most probably destroying the
entanglement of the resulting states. To lower them we can search
for symmetric product vectors $\ket{e}^{\ot N}\in H_{2,N}$ such
that their respective partial conjugations belong to all the
ranges of the relevant partial transpositions
$R(\omega_{N,\lambda}^{T_i})$ $(i=A,AB,\ldots,)$. Every such
vector can be removed from $\omega_{N,\lambda}$, i.e., we consider
a state $\widetilde{\omega}_{N,\lambda}=\omega_{N,\lambda}-\mu
P_{\ket{e}^{\ot N}}$, where $P_{\ket{e}^{\ot N}}$ denotes a
projector onto $\ket{e}^{\ot N}$. By properly choosing $\mu$ and
the product vector $\ket{e}^{\ot N}$ provided they exist, we can
lower some of the ranks of partial transpositions of
$\omega_{N,\lambda}$.

Let us now follow the above general recipe and get the
aforementioned four-qubit symmetric PPT entangled states. To this
end, we take $\rho'_{5,b}=(\mathbbm{1}\ot
F)\rho_{5,b}(\mathbbm{1}\ot F^{\dagger})$ [cf. Eq.
(\ref{states})], and, following the above description, apply the
local filter $F_2$, which is now $4\times 5$ matrix with two
parameters $\gamma_1$ and $\gamma_2$, and subsequently, the next
filter $V$. This results in a family of four-qubit symmetric
states $\omega_{4,b,\gamma_1,\gamma_2}$ parameterized by $b$ and
$\gamma_i$ $(i=1,2)$. In order to get a particular example
four-qubit PPTESS, let us put $b=1/2$ and
$\gamma_1=\gamma_2^{-1}=1/\sqrt{2}$, which leads us to the state
$\omega_{4}$ such that $\omega_{4}^{T_A}\geq 0$, while
$\omega_{4}^{T_{AB}}\ngeq 0$. To ``cover" the negative eigenvalues
of $\omega_{4}^{T_{AB}}$, we consider
${\omega}_{4,\lambda}=\omega_{4}+\lambda P_4$. With the aid of
numerics one finds that $\lambda_*\thickapprox0.94842$ is the
smallest $\lambda$ for which
${\omega}_{4,\lambda_{*}}^{T_{AB}}\geq 0$. However, the three-rank
of ${\omega}_{4,\lambda_{*}}$ is $(5,8,8)$. We can then lower the
second rank by subtracting a product vector $\ket{e}^{\ot 4}\in
R(\omega_{4,\lambda_*})$ such that $\ket{e^{*}}\ket{e}^{\ot 3}\in
R(\omega_{4,\lambda_*}^{T_A})$ and $\ket{e^{*},e^{*},e,e}\in
R(\omega_{4,\lambda_*}^{T_{AB}})$. Again, exploiting numerics, we
find that there exists such a vector $\ket{e}=(1,\alpha)$, where
$\alpha\thickapprox7 + 38.52091\mathrm{i}$. Then, one checks that
for $\mu\thickapprox 0.64625$, $\omega_{4,\lambda_*}-\mu
P_{\ket{e}^{\ot 4}}$ after normalization is the expected example
of four-qubit symmetric PPT entangled state with three-rank
$(5,7,8)$. Using the algorithm below one can check that the state
is extremal, and thus also edge. It should be noticed that the
above choice of parameters $\gamma_i,b$ was made for simplicity,
but other choices can also lead to PPT entangled states (e.g.,
$\gamma_1=3/8$, $\gamma_2=11/23$, $b=1/6$).


{\it Searching for extremal PPT entangled four-qubit symmetric
states.--} We have just shown that four-qubit symmetric PPT
entangled states exist. Clearly, there must then exist extremal
entangled elements in the corresponding set of PPT states
$D_{\mathrm{PPT}}^{\mathrm{sym}}$. Our aim now is to search for
such states and characterize them. For this purpose we adapt the
algorithm for searching of extremal elements in the set of PPT
states, originally proposed for bipartite systems \cite{Leinaas}
(see also Ref. \cite{OptComm}), to the multipartite scenario.
Then, we will apply it to $D_{\mathrm{PPT}}^{\mathrm{sym}}$.

Let us consider again the Hilbert space
$H_{d,N}=(\mathbbm{C}^{d})^{\ot N}$ and the set $D_{\mathrm{PPT}}$
of all PPT states acting on $H_{d,N}$. Let $\rho\in
D_{\mathrm{PPT}}$ and let $\mathcal{P}$ and $\mathcal{P}_k$ denote
projectors onto, respectively, $R(\varrho)$ and $R(\rho^{T_k})$
$(k=1,\ldots,M)$, where $M$ denotes the number of independent
partial transpositions (recall that some of them are equivalent
under the full transposition).

The state $\rho$ is extremal in $D_{\mathrm{PPT}}$ iff it cannot
be written as $\rho=p\rho_1+(1-p)\rho_2$ for some $\rho_i\in
D_{\mathrm{PPT}}$ and $0<p<1$. This is equivalent to say that
$\rho$ is not extremal iff there exists a density matrix
$\sigma\neq \rho$ such that $R(\sigma)\subseteq R(\rho)$ and
$R(\sigma^{T_k})\subseteq R(\rho^{T_k})$ for all $k$. One can even
relax the assumption of $\sigma$ being positive to be Hermitian
and such Hermitian matrices are solutions to system of equations
\begin{equation}\label{system}
\hat{\mathcal{P}}_{k}(h)=h \qquad (k=0,\ldots,M),
\end{equation}
where $\hat{\mathcal{P}}_0=\mathcal{P}(\cdot)\mathcal{P}$ and
$\hat{\mathcal{P}}_{k}(\cdot)=[\mathcal{P}_{k}(\cdot)^{T_k}\mathcal{P}_{k}]^{T_k}$
$(k=1,\ldots,M)$. This system is equivalent to the single equation
$[\hat{\mathcal{P}}_M\circ\ldots
\circ\hat{\mathcal{P}}_1\circ\hat{\mathcal{P}}](h)=h$. Clearly,
$\rho$ is a particular solution of the system (\ref{system}) and,
due to the above statements, is extremal iff it is its only
solution, which gives necessary and sufficient condition for
extremality \cite{Leinaas,OptComm}. This also leads to a simple
necessary criterion for extremality. Precisely, each equation in
(\ref{system}) imposes $d^{2N}-[r(\rho^{T_k})]^{2}$ linear
constraints on the matrix $h$. The maximal number of conditions
imposed by the system (\ref{system}) is then
$\sum_{k}(d^{2N}-[r(\rho^{T_k})]^{2})$. Then, a Hermitian matrix
$h$ has $d^{2N}$ real parameters, and hence if
\begin{equation}\label{cond}
\sum_{k=0}^{M}[r(\rho^{T_k})]^{2}\geq Md^{2N}+1,
\end{equation}
the state $\rho$ is not extremal.

All this induces a method of searching for the extremal states in
$D_{\mathrm{PPT}}$ \cite{Leinaas}. Taking $\rho\in
D_{\mathrm{PPT}}$, one solves the corresponding system
(\ref{system}). If the latter has a solution $h\neq \rho$, the
state $\rho$ is not extremal. One then considers a family of
matrices $\rho(x)=(1-x\Tr h)\rho+xh$ with $x$ being in general
a real parameter. Clearly, there is $x=x_{*}$ such that
$\rho_2=\rho(x_{*})$ is a still PPT state, however, either
$r(\rho_2)=r(\rho)-1$ or $r(\rho_2^{T_k})=r(\rho^T_{k})-1$ for
some $k$. We can again apply the above procedure to $\rho_2$, and
in case it is not extremal get another PPT state $\rho_3$ with at
least one of the ranks diminished by one. We keep applying this
procedure until we obtain an extremal state, which appears in a
finite number of repetitions. If the resulting state is pure, it
is separable, otherwise it is entangled. Notice that in order to
get a particular extremal entangled state one has to properly
choose the initial state which basically means that it should be
of higher ranks, as for instance the maximally mixed state, and
the directions which follow from solving Eqs. (\ref{system}).

Let us apply the above algorithm to the symmetric states. In this
case the left-hand side of (\ref{cond}) has to be modified as
$\rho$ and $\rho^{T_X}$ $(X=A,AB,\ldots)$ are supported on Hilbert
spaces of different dimensions. In particular, for $N=4$ such
analysis was done in Ref. \cite{OptComm} and it follows that
states of ranks $(5,7,9)$, $(5,8,8)$ cannot be extremal. Moreover,
theorems (\ref{thm_sep}) and \ref{thm_edge} imply that generic
states of ranks $(5,8,7)$ and $(5,r(\rho^{T_A}),r(\rho^{T_{AB}}))$
with $r(\rho^{T_A})$ or $r(\rho^{T_{AB}})\leq 6$ also cannot be
extremal. The natural candidates for extremal states have then
ranks $(5,7,7)$ and $(5,7,8)$.

We applied the above algorithm to four-qubit PPT states and all
the extremal examples we found have ranks $(5,7,8)$. In 30 000
runs we generated 5760 unitarily nonequivalent extremal entangled
states and all of them have ranks $(5,7,8)$. As an initial state
we took the projector $P_4$ onto $S_4$ (recall that the initial
state has to be of rank five and due to theorem \ref{thm_sep} must
also have appropriately high ranks of $\rho^{T_X}$ $(X=A,AB)$). At
each step of the algorithm we used solutions of (\ref{system})
chosen so that we could reach one of the three-ranks not excluded
by theorem \ref{thm_edge}. We also got 24 240 states of ranks
$(5,7,7)$ in this way but they all are separable.

{\it Conclusion.--} The main aim of this note was to solve the
open question of the existence of four-qubit PPT entangled
symmetric states. We have reached this goal by proposing a
half-analytical-half-numerical method of constructing of such
states. The analytical part of the method allows one to map a
class of qubit-qudit PPT entangled states onto many-qubit
entangled symmetric states. Then, using already-well-established
methods of the theory of entanglement, and with the help of
numerics, we have found the desired PPT entangled states.

We have also characterized the four-qubit PPT symmetric states
with respect to separability, edgeness and extremality properties.
First, we have proven that generic states of a given three-rank
different from $(5,7,7)$ or $(5,7,8)$ are not edge. Then, by
adapting to the multipartite case an algorithm allowing to search
for extremal PPT states \cite{Leinaas}, we have sought extremal
four-qubit PPT entangled symmetric states. All the entangled
states found in this way have ranks $(5,7,8)$, while those with
ranks $(5,7,7)$ encountered in this way are separable (see Ref.
\cite{Jordi} for more details).

Interestingly, all the methods presented in this paper can be
applied to $N$-qubit symmetric states. For instance, we have shown
that with our method one can obtain five-qubit and six-qubit PPT
entangled symmetric states, confirming the findings of Refs.
\cite{Perm1,Perm2}. Generalization of these findings to $N$-qubit
symmetric Hilbert spaces is currently being studied and will be a
subject of a forthcoming publication \cite{Jordi}.


{\it Acknowledgments.--} This work is supported by EU project
AQUTE, ERC Grants QUAGATUA and QOLAPS, ERC Starting Grant
GEDENTQOPT, EU projects CHIST-ERA DIQIP and QUASAR, Spanish MINCIN
project FIS2008-00784 (TOQATA), and Polish MNiSW grant "Ideas
Plus" no. IdP2011 000361. R. A. acknowledges the support from the
Spanish MINCIN through the Juan de la Cierva program.


\appendix

\section{Characterization of four-qubit symmetric states}
\label{appA}

Here we recall theorems \ref{thm_sep}, \ref{thm_edge}, and
\ref{thm_dec} and prove them in detail.

\begin{thmm}\label{thm_sep2}
Let $\rho$ be a four-qubit PPT symmetric state. If either
$r(\rho)\leq 4$, or $r(\rho^{T_A})\leq 4$, or
$r(\rho^{T_{AB}})\leq 3$ then $\rho$ is separable, while if
$r(\rho^{T_A})\leq 6$, or $r(\rho^{T_{AB}})\leq 6$, then
generic $\rho$ is separable.
\end{thmm}

\begin{proof}Although the first two cases of $r(\rho)\leq 4$
and $r(\rho^{T_A})\leq 4$ were already proven in Ref.
\cite{EckertAnP} let us, for completeness, recall their proofs.
First, due to the fact that $S_3\cong\mathbbm{C}^4$, one can
always treat $\rho$ as a qubit-ququart PPT state. Then, the
results of Ref. \cite{2N} say that any qubit-ququart PPT state of
rank $r(\rho)\leq 4$ is separable. Replacing then $\rho$ by
$\rho^{T_A}$ and following the same arguments, one proves the case
of $r(\rho^{T_A})\leq 4$.

In order to prove the case of $r(\rho^{T_{AB}})\leq 3$, one
considers a state $\sigma=\rho^{T_{AB}}$ and exploits the fact
that $S_2$ is isomorphic to $\mathbbm{C}^3$. Then, $\sigma$ is a
two-qutrit PPT state such that $r(\sigma)\leq 3$, and it was shown
in Ref. \cite{MN} that any two-qutrit PPT state of rank less or
equal to three is separable.

In the case of $r(\rho^{T_A})\leq 6$ let us define
$\sigma=\rho^{T_A}$ and consider it as a bipartite state with
respect to the partition $B|ACD$. Clearly, in such case $\sigma$
acts on $\mathbbm{C}^2\ot\mathbbm{C}^6$ and is of rank at most
six. Provided that it is supported on
$\mathbbm{C}^2\ot\mathbbm{C}^6$, which generically is the case,
the results of Ref. \cite{2N} tell us that $\sigma$ is separable
across $B|ACD$, i.e.,
\begin{equation}\label{dec1}
\sigma=\rho^{T_A}=\sum_{i}p_{i}\proj{e_i}_B\ot\proj{\psi_i}_{ACD},
\end{equation}
where, for the time being, $\ket{\psi_i}$ are entangled states
from $\mathbbm{C}^2\ot S_2$. On the other hand, the $BCD$
subsystem of $\sigma$ is still supported on the three-qubit
symmetric subspace. As a result, any vector
$\ket{e_i}\ket{\psi_i}$ in the decomposition (\ref{dec1}) must
obey
$P_3\ket{e_i}_{B}\ket{\psi_i}_{ACD}=\ket{e_i}_{B}\ket{\psi_i}_{ACD}$,
where $P_3$ is applied to $BCD$ subsystem. This, after some
algebra, implies that
$\ket{\psi_i}_{ACD}=\ket{f_i}_A\ket{e_i}_C\ket{e_i}_D$ for some
$\ket{f_i}\in\mathbbm{C}^2$, and hence $\sigma$, and accordingly
$\rho=\sigma^{T_A}$ are fully separable.

To prove the last case of $r(\rho^{T_{AB}})\leq 6$ one follows the
same lines as before substituting $\rho^{T_{AB}}$ for
$\rho^{T_A}$.
\end{proof}

\begin{thmm}\label{thm_edge2}
Generic four-qubit PPT symmetric states of a given three-rank
different from $(5,7,7)$ and $(5,7,8)$ are not edge.
\end{thmm}
\begin{proof}We will show that in all relevant
cases with respect to the three-rank, except for $(5,7,7)$ and
$(5,7,8)$, there exists a symmetric product vector $\ket{e}^{\ot
4}\in R(\rho)$ such that $\ket{e^{*}}\ket{e}^{\ot 3}\in
R(\rho^{T_A})$, and $\ket{e^{*},e^{*},e,e}\in R(\rho^{T_{AB}})$.
Clearly, many of the three-ranks can be ruled out with the aid of
theorem \ref{thm_sep}, and the remaining ones are $(5,7,7)$,
$(5,7,8)$, $(5,8,7)$, $(5,8,8)$, $(5,7,9)$, and $(5,8,9)$. The
last one is trivial because all symmetric product vectors
$\ket{e}^{\ot 4}$ belong to $R(\rho)$ and their respective partial
conjugations to $R(\rho^{T_A})$ and $R(\rho^{T_{AB}})$. In what
follows we give a proof for the cases $(5,8,7)$, $(5,8,8)$,
$(5,7,9)$.

In the case of $\widetilde{r}(\rho)=(5,7,9)$, $\rho$ and
$\rho^{T_{AB}}$ are of full rank and therefore one has to find a
product vector $\ket{e^*,e,e,e}$ which is orthogonal to the only
vector $\ket{\Psi}$ from $K(\rho^{T_A})$. To this end, let us
write $\ket{\Psi}=\ket{0}\ket{\psi_0}+\ket{1}\ket{\psi_1}$ with
$\ket{\psi_i}\in S_3$, and take $\ket{e}=(1,\alpha)$
$(\alpha\in\mathbbm{C})$. The orthogonality condition then reads
$V_3(\alpha)+\alpha^{*}W_3(\alpha)=0$ with $V_3$ and $W_3$
denoting polynomials of degree at most three over the complex
field. In Ref. \cite{Normal}, this equation was shown to have
generically (both polynomials $V_3$ and $W_3$ are of degree three)
at least one solution. Consequently, generic four-qubit PPT states
of of ranks $(5,7,9)$ are not edge.

In the case of $\widetilde{r}(\rho)=(5,8,8)$, $\rho$ and
$\rho^{T_A}$ are of full rank, and so one has to find a product
vector $\ket{e^*e^*ee}$ with $\ket{e}\in\mathbbm{C}^2$ orthogonal
to the only vector $\ket{\Psi}$ from the kernel of $\rho^T_{AB}$.
For this purpose, let us notice that
$\rho^{T_{AB}}=G(\rho^{*})^{T_{AB}}G^{\dagger}$, where $G$ denotes
an operator swapping subsystems $AB$ and $CD$. This means that
$\ket{\Psi}$ enjoys the same symmetry, i.e.,
$G\ket{\Psi}=\ket{\Psi^{*}}$. As a result, one can express it as
\begin{equation}\label{dec}
\ket{\Psi}=\sum_{k=1}^{3}\lambda_{k}\ket{e^*_k}_{AB}\ket{e_k}_{CD},
\end{equation}
where $\lambda_{k}\in\mathbbm{R}$ and $\ket{e_k}$ are orthogonal
symmetric two-qubit vectors. Exploiting the fact that
$\ket{\Psi}\in K(\rho^{T_{AB}})$, one sees that
\begin{eqnarray}
\sum_{k=1}^{K}\lambda_k\bra{x^*,y}\rho^{T_{AB}}\ket{e^*_k,e_k}&=&
\sum_{k=1}^{K}\lambda_k\bra{e_k,y}\rho\ket{x,e_k}\nonumber\\
&=&\sum_{k=1}^{K}\lambda_k\bra{e_k,x}\rho\ket{e_k,y}=0\nonumber\\
\end{eqnarray}
holds for any pair of qubit vectors $\ket{x}$ and $\ket{y}$ with
the second equality stemming from the fact that $\rho$ is
symmetric and hence $\rho=\rho G$.
This immediately implies that
\begin{equation}\label{rownanie}
\sum_{k}\lambda_{k}\langle e_k|\rho|e_k\rangle=0,
\end{equation}
where the right-hand side is a two-qubit matrix acting on the $CD$
subspace, obtained by "sandwiching" $\rho$ with $\ket{e_k}$s on
the first two qubits.

On the other hand, taking into account Eq. (\ref{dec}), there
exists $\ket{e}\in\mathbbm{C}^2$ such that $\ket{e^*e^*ee}\in
R(\rho^{T_{AB}})$ iff
\begin{equation}\label{expr}
\langle e,e|\left[\sum_{k}\lambda_k
\proj{e_k}\right]|e,e\rangle=0.
\end{equation}
In order to show that such $\ket{e}$ exists, assume, in contrary,
that Eq. (\ref{expr}) does not hold for any
$\ket{e}\in\mathbbm{C}$. Then, its left-hand side must have the
same sign for all $\ket{e}$, say positive (as otherwise, from
continuity, there would exist $\ket{e}$ for which (\ref{expr})
holds). Consequently,
\begin{equation}
\langle e,e|\left[\sum_{k}\lambda_k \proj{e_k}\right]|e,e\rangle
>0
\end{equation}
for any $\ket{e}\in\mathbbm{C}^2$, which, owing to the fact that
$\ket{e_k}$ are symmetric, implies that $W=\sum_{k}\lambda_k
\proj{e_k}$ is a two-qubit entanglement witness. Since all
two-qubit witnesses are decomposable, we have $W=P+Q^{T_{A}}$ with
$P,Q\geq 0$. This, when substituted to Eq. (\ref{rownanie}),
implies that the two conditions
\begin{equation}
\Tr[(P\ot \ket{y}\!\bra{x})\rho]=0,\qquad\Tr[(Q\ot
\ket{y}\!\bra{x})\rho^{T_A}]=0
\end{equation}
%
%
must be obeyed for any $\ket{x},\ket{y}\in\mathbbm{C}^2$,
contradicting the fact that $\rho$ and $\rho^{T_A}$ are of full
rank. Notice that this proof is general (not generic) meaning that
there are no four-qubit symmetric edge states of ranks $(5,8,8)$.

Let us now pass to the most involving case of
$\widetilde{r}(\rho)=(5,8,7)$. Here $\rho$ and $\rho^{T_A}$ are of
full rank, while $K(\rho^{T_{AB}})$ has dimension two. Hence, to
find a product vector $\ket{e}^{\ot 4}\in R(\rho)$ such that
$\ket{e^*,e^*,e,e}\in R(\rho^{T_{AB}})$, one has to solve two
equations $\langle{e^*,e^*,e,e}|\Psi_i\rangle=0$ $(i=1,2)$, where
$\ket{\Psi_i}$ are two orthogonal vectors from $K(\rho^{T_{AB}})$.
Exploiting again the identity
$\rho^{T_{AB}}=G(\rho^{*})^{T_{AB}}G^{\dagger}$, it is fairly easy
to see that they can be written as
\begin{equation}\label{wektory}
\ket{\Psi_1}=\sum_{k=1}^{2}\lambda_{k}\ket{e_k}\ket{f_k^{*}},\qquad
\ket{\Psi_2}=\sum_{k=1}^{2}\lambda_{k}\ket{f_k}\ket{e_k^{*}}.
\end{equation}
To see it explicitly, let us first notice that we can assume that
one of $\ket{\Psi_i}$ is of Schmidt rank two. If both of them are
of rank three, then there exists a vector of Schmidt rank two in
$\mathrm{span}\{\ket{\Psi_1},\ket{\Psi_2}\}$. On the other hand,
if one of $\ket{\Psi_i}$ $(i=1,2)$ is of rank one, i.e., is
product with respect to the partition $AB|CD$, $r(\rho)=4$
contradicting the assumption that $\rho$ is entangled. Assume then
$\ket{\Psi_1}$ is of rank two. Then either $G\ket{\Psi_1^{*}}$ is
linearly independent of $\ket{\Psi_1}$ leading to Eq.
(\ref{wektory}), or $G\ket{\Psi_1^*}=\xi\ket{\Psi_1}$ for some
$\xi\in\mathbbm{C}$. In the latter case, short algebra implies
that $\ket{\Psi_i}$ $(i=1,2)$ are not linearly independent
contradicting the fact that they span two-dimensional kernel of
$\rho^{T_{AB}}$.

As a result, finding a vector $\ket{e^*,e^*,e,e}\in
R(\rho^{T_{AB}})$ is equivalent to solving an equation
\begin{equation}\label{equation0}
V(\alpha^{*})W(\alpha)+\widetilde{V}(\alpha^{*})\widetilde{W}(\alpha)=0,
\end{equation}
where $V,\widetilde{V}$ and $W,\widetilde{W}$ are polynomials
generically of degree two. A solution to this equation exists if
and only if there exists $z\in\mathbbm{C}$ such that
\begin{equation}
V(\alpha^{*})=z\widetilde{V}(\alpha^{*})
\end{equation}
and
\begin{equation}
\widetilde{W}(\alpha)=-zW(\alpha).
\end{equation}
We have then brought a single equation of the fourth degree to two
equations of the second degree. With the aid of the first one, we
can determine $\alpha^*$ as a function of $z$. There are clearly
at most two such solutions. Putting them to the second equation
and getting rid of the square root, we arrive at
\begin{equation}\label{equation}
(z^{*})^{2}W_{4}(z)+z^{*}W'_4(z)+W''_4(z)=0,
\end{equation}
where $W_4$, $W_4'$, and $W_4''$ stand for polynomials which are
generically of fourth degree. In what follows, we will show that
Eq. (\ref{equation}) has at least one solution $z=rs$ with
$|s|=1$, i.e., $z^{*}=r/s$. To this end, let us consider two
cases, when $s=rx$ and $s=x/r$. Putting all this to Eq.
(\ref{equation}), one gets the following equations
\begin{equation}\label{Eq1}
\left(\frac{1}{x}\right)^2W_4(r^2x)+\frac{1}{x}W_4'(r^2x)+W''_4(r^2x)=0
\end{equation}
and
\begin{equation}\label{Eq2}
\left(\frac{r^2}{x}\right)^{2}W_4(x)+\frac{r^2}{x}W'_4(x)+W''_4(x)=0.
\end{equation}
In the limit of $r\to\infty$ the first equation has two roots
$s_i^{\infty}\to \infty$ $(i=1,2)$, while the second one four
roots $s_i^{\infty}\to 0 $ $(i=1,2,3,4)$. Then, in the limit of
$r\to 0$, Eq. (\ref{Eq1}) again has two roots $s_i^{0}\to 0$,
while Eq. (\ref{Eq2}) has four roots $s_i^{0}\to \infty$
$(i=1,2,3,4)$. Notice that, after the above substitution,
(\ref{equation}) is of sixth degree in $s$ meaning that it can
have at most six solutions with respect to $s$. Consequently, by
varying continuously $r$ from zero to $\infty$ we see that at
least one of the roots $s_i$, being a continuous function of $r$,
must go from zero to $\infty$, and so there is a value of $s$ such
that $|s|=1$. As a result, there is at least one $z\in\mathbbm{C}$
for which Eq. (\ref{equation}) is fulfilled, and simultaneously at
least one $\alpha\in\mathbbm{C}$ obeying (\ref{equation0}).
 \end{proof}

 \begin{thmm}\label{thm_dec2}
Let $\rho$ be an entangled symmetric PPT four-qubit state. Then
it can be written as
\begin{equation}\label{Lemma12}
\rho=\sum_{k=1}^{K\leq 6} [A_k (a_k,b_k)^{\ot 4}+B_k
(a_k,-b_k)^{\ot 4}],
\end{equation}
where $(a_k,b_k)\in\mathbbm{C}^{2}$ and $A_k,B_k$ are some complex
coefficients, and by $[\psi]$ we denote a projector onto
$\ket{\psi}$.
\end{thmm}

\begin{proof} First, let us introduce two linear transformations
$V:(\mathbbm{C}^2)^{\ot 2}\mapsto \mathbbm{C}^2$ and
$W:(\mathbbm{C}^{2})^{\ot 2}\mapsto (\mathbbm{C}^{2})^{\ot 3}$
defined as
\begin{equation}
V[(a,b)\ot(a,b)]=(a^2,b^2)
\end{equation}
and
\begin{equation}
W[(a,b)\ot(a^2,b^2)]=(a,b)^{\ot 3},
\end{equation}
respectively, with $a,b$ being any complex numbers. Then, by
$\hat{V}$ and $\hat{W}$ we denote maps that are defined through
the adjoint actions of $V$ and $W$, i.e.,
$\hat{X}(\cdot)=X(\cdot)X^{\dagger}$ $(X=V,W)$.

The key feature of the two matrices $V$ and $W$ is that
$WV_{BC}\ket{\psi}=\ket{\psi}$ for any three-qubit symmetric
$\ket{\psi}$, where $BC$ denote qubits subject to $V$. The same
holds when $V$ is applied to any pair of qubits in $\ket{\psi}$
and followed by a proper application of $W$. Accordingly, any
$N$-qubit symmetric state $\rho$ is left invariant under a
proper application of $\hat{V}$ and $\hat{W}$ to any three-qubits.
In particular, for a four-qubit symmetric state $\rho$,
$\hat{W}_{BC}\circ\hat{V}_{CD}(\rho)=\rho$.

Let us now consider a four-qubit PPT symmetric state $\rho$. By
applying $\hat{V}$ to the $CD$ subsystem of $\rho$, we get a
three-qubit state $\sigma_{ABC'}=\hat{V}_{CD}(\rho)$ acting on
$S_2\ot\mathbbm{C}^2$, where $C'$ denotes the qubit resulting from
the application of $\hat{V}$. Clearly, the map $\hat{V}$ preserves
positivity of partial transposition with respect to the first two
parties, i.e., $\sigma^{T_{AB}}\geq 0$. Since $S_2$ is isomorphic
to $\mathbbm{C}^3$, results of Ref. \cite{Horodecki96PLA} imply
that $\sigma$ is separable across the cut $AB|C'$ and so $\sigma$
takes the form
\begin{equation}\label{decomposition}
\sigma=\sum_{k=1}^{K}\proj{e_k}\ot \proj{f_k},
\end{equation}
with $\ket{e_k}\in\mathbbm{C}^{3}$ and $\ket{f_k}\in\mathbbm{C}^2$
being in general unnormalized vectors from $\mathbbm{C}^3$ and
$\mathbbm{C}^2$, respectively, and $K\leq 6$ \cite{Normal}.

By the very assumption $\rho$ is entangled and therefore
$r(\rho)=5$, which together with the fact that $r(V)=2$, mean
that the rank of $\sigma$ is also five. Therefore, $K(\sigma)$
consists of a single vector
$\ket{\phi}\in\mathbbm{C}^3\ot\mathbbm{C}^2$, which, due to the
fact that the range of $\sigma$ is spanned by the vectors
$(a,b)^{\ot 2}\ot (a^{2},b^{2})$, takes the form
$\ket{\phi}=\ket{01}-\ket{20}$. As a result, any product vector in
Eq. (\ref{decomposition}) has to be orthogonal to $\ket{\phi}$.

Putting $\ket{f_k}=(a_k^2,b_k^2)$ with $a_k,b_k\in\mathbbm{C}$ and
solving the equation $\langle\phi|e_k,f_k\rangle=0$ with respect
to $\ket{e_k}$ one finds that it can be written as
$\ket{e_k}=A_k(a_k^2,a_k b_k,b^2_k)+B_k(a_k^2,-a_k b_k,b^2_k)$
with some $A_k,B_k\in\mathbbm{C}$. Putting the above forms of
$\ket{e_k}$ and $\ket{f_k}$ to Eq. (\ref{decomposition}), one sees
that $\sigma$ can be written as
\begin{equation}
\sigma=\sum_{k=1}^{K}[(A_k (a_k,b_k)^{\ot 2}+B_k(a_k,-b_k)^{\ot
2})\ot(a_k^{2},b_k^{2})],
\end{equation}
where $K\leq 6$ and $[\psi]$ denotes a projector onto
$\ket{\psi}$. One completes the proof by applying $\hat{W}$ to the
last two qubits of $\sigma$.
\end{proof}

Utilizing the normal matrix approach to the separability problem
\cite{Normal}, one can prove a bit different decomposition.

\begin{thmm}\label{thm_dec22}
Let $\rho$ be an entangled symmetric PPT four-qubit state. Then
it can be written as
\begin{equation}\label{Lemma13}
\rho=\sum_{k=1}^{K} [A_k (a_k,b_k)^{\ot 4}+B_k (0,1)^{\ot 4}],
\end{equation}
where $K\leq 6$, $(a_k,b_k)\in\mathbbm{C}^{2}$, and $A_k,B_k$ are
some complex coefficients, and by $[\psi]$ we denote a projector
onto $\ket{\psi}$.
\end{thmm}
\begin{proof}
The proof exploits the method developed in Ref. \cite{Normal}.
First, one notices that any $\rho$ can be written as a sum of
rank-one matrices
\begin{equation}\label{ensemble}
\rho=\sum_{i=1}^{K}\proj{\Psi_i},
\end{equation}
where, in particular, $\ket{\Psi_i}$ can be (unnormalized)
eigenvectors of $\rho$, and $K\leq 6$ (see the proof of theorem
\ref{thm_dec2}). On the other hand, $\rho$ can always be
expressed with the aid of the symmetric unnormalized basis
$\{\ket{E^4_{\mu}}\}_{\mu=1}^{5}$ spanning
$S_4$ as
\begin{equation}\label{decom2}
\rho=\sum_{\mu,\nu=1}^{5}\rho_{\mu\nu}\ket{E^4_{\mu}}\!\bra{E^4_{\nu}}.
\end{equation}
Both decompositions (\ref{ensemble}) and (\ref{decom2}) are
related {\it via} the so-called Gram system of $\rho$, i.e., a
collection of $K$-dimensional vectors $\ket{v_{\mu}} = (1/\langle
E^4_{\mu}|E^4_{\mu}\rangle)(\langle
\Psi_1|E^4_{\mu}\rangle,\ldots,\langle \Psi_K|E^4_{\mu}\rangle)$
$(\mu=1,\ldots,5)$, giving $\rho_{\mu\nu}=\langle
v_{\mu}|v_{\nu}\rangle$. Putting the latter to (\ref{decom2}) with
explicit forms of the vectors $\ket{v_{\mu}}$, one recovers
(\ref{ensemble}).

Now, by projecting the last party onto $\ket{0}$ we get a
three-qubit symmetric PPT state $\widetilde{\rho}$, which, as
already stated,  is separable. Then, according to Ref.
\cite{Normal}, there exists a diagonal matrix
$M=\mathrm{diag}[\alpha_1^{*},\ldots,\alpha_K^{*}]$ such that
$\ket{v_{\mu}}=M^{\mu-1}\ket{v_{1}}$ $(\mu=1,\ldots,4)$. For
convenience we can also put
$\ket{v_{5}}=M^{4}\ket{v_1}+\ket{\widetilde{v}}$ with
$\ket{\widetilde{v}}$ being some $K$-dimensional vector. Then,
putting $\ket{v_1}=(A_1^{*},\ldots,A_K^{*})$ and
$\ket{\widetilde{v}}=(B_1^{*},\ldots,B_K^{*})$, one sees that
\begin{eqnarray}
\ket{\Psi_i}&=&\sum_{\mu=1}^{5}\frac{\langle E^4_{\mu}|\Psi_i\rangle}{\langle E^4_{\mu}|E_{\mu}\rangle}\ket{E^4_{\mu}}\nonumber\\
&=&A_i\sum_{\mu=1}^{4}\alpha_{i}^{\mu-1}\ket{E^4_k}+B_i\ket{E^4_5}\nonumber\\
&=&A_i(1,\alpha_i)^{\ot 4}+B_i\ket{E^4_5},
\end{eqnarray}
where the second equation follows from the explicit form of the
vectors $\ket{v_{\mu}}$. Substituting vectors $\ket{\Psi_i}$ to
Eq. (\ref{ensemble}), one gets (\ref{Lemma13}), which completes the
proof.
\end{proof}

\section{Properties of the states $\rho_{d,b}$}
\label{appB}

Here we characterize the states (\ref{states}) in more details. In
particular, we prove that for $d\geq 4$ and $b\in[0,1]$ they are
PPT entangled and edge.

\begin{thmm}\label{thm_PPT}
The states $\rho_{d,b}$ are PPT for $d\geq 2$ and $b\in[0,1]$.
\end{thmm}
\begin{proof}
First, notice that we can rewrite (\ref{states}) in the matrix
form as
\begin{equation}
    \rho_{d,b}=\frac{1}{(2d-1)b+1}\left(
    \begin{array}{cc}
        C & b B_U\\
        b B_L & b\Eins_d
    \end{array}\right),
\end{equation}
where $B_U$ and $B_L$ are $d\times d$ dimensional matrices with
entries 1 on the upper and lower diagonal, respectively. Further,
$C$ is a $d\times d$ matrix given by
\begin{equation}
    C=\frac{1}{2}\left(\begin{array}{ccccc}
         1+b & 0 &  \cdots  & 0 &\sqrt{1-b^2}\\
         0 & b & \cdots & 0 & 0\\
         \vdots & \vdots & \ddots & \vdots &  \vdots\\
         0 & 0 & \cdots & b & 0\\
         \sqrt{1-b^2} & 0 & \cdots  & 0 & 1+b
    \end{array}\right).
\end{equation}
Clearly, the partial transposition of $\rho_{b,d}$ reads
\begin{equation}
    \rho_{d,b}^{T_A}=\frac{1}{(2d-1)b+1}\left(
    \begin{array}{cc}
        C & b B_L\\
        b B_U & b\Eins_d
    \end{array}\right).
\end{equation}
Now, let us consider the unitary matrix
$U=\mathrm{antidiag}[1,\ldots,1]$  (anti-diagonal matrix
consisting of unities). Straightforward calculations show that $U
B_U U^{\dagger}=B_L$, $U B_L U^{\dagger}=B_U$, and
$UCU^{\dagger}=C$. Consequently,
\begin{equation}
\rho_{b,d}^{T_A}=(\mathbbm{1}_2\ot U)\rho_{b,d}(\mathbbm{1}_2\ot
U^{\dagger}),
\end{equation}
meaning that $\rho_{b,d}^{T_A}\geq 0$ iff $\rho_{b,d}\geq 0$.
\end{proof}

\begin{thmm}
The states $\rho_{d,b}$ are entangled for $d\geq4$ and
$b\in(0,1)$, while separable for $d=2,3$, or $b=0$, or $b=1$.
\end{thmm}
\begin{proof}
First, let us prove that for $d\geq 4$ and $b\in(0,1)$, the states
(\ref{states}) are entangled. For this purpose, it suffices to use
the necessary criterion for separability formulated in Ref.
\cite{Horodecki97PLA} -- the range criterion. It says that if a
given density matrix $\rho$ is separable then one is able to
find product vectors $\ket{e,f}$ spanning $R(\rho)$ such that
$\ket{e^*,f}$ span $R(\rho^{T_A})$. In what follows we show
that none of the product vectors $\ket{e,f}$ in $R(\rho)$ is
such that $\ket{e^*,f}\in R(\rho^{T_A})$.

All product vectors in the range of $\rho_{b,d}$ are given by
\begin{eqnarray}
\label{Prvecs1}(1,\alpha)&\ot&
(\alpha^{d-1}+y,\alpha^{d-2},\ldots,\alpha,1) \qquad
(\alpha\in\mathbbm{C})\\ \label{Prvecs2}(0,1)&\ot& (1,0\ldots,0),
\end{eqnarray}
where $y=\sqrt{(1-b)/(1+b)}$. If we allow for infinite $\alpha$,
the vector (\ref{Prvecs2}) may be obtained from the class
(\ref{Prvecs1}). It is also worth mentioning that the above
vectors span $R(\rho_{d,b})$.

On the other hand, all the vectors in the range of
$\rho_{b,d}^{T_A}$ are given by
\begin{equation}\label{range2}
(a_1,\ldots,a_{d-1},ya_1+a_d;a_2,a_3,\ldots,a_d,b)
\end{equation}
with $a_1,\ldots,a_d,b\in\mathbbm{C}$. Consequently, a product
vector from the first class (\ref{Prvecs1}), when partially
conjugated with respect to the first subsystem, belongs to
$R(\rho_{b,d}^{T_A})$, i.e., takes the form (\ref{range2}), if and
only if the conditions are satisfied: (i)
$\alpha(1-|\alpha|^2)=0$, (ii)
$\alpha^{d-2}=\alpha^{*}(\alpha^{d-1}+y)$, and (iii)
$y(y+\alpha^{d-1})=1+|\alpha|^{2}$. The first condition is
satisfied if either $\alpha=0$, which contradicts the third
condition because $y\neq 1$, or $|\beta|^2=1$, which contradicts
(ii) because $y\neq 0$. Along the same lines one checks that the
vector (\ref{Prvecs2}) is not of the form (\ref{range2}).

In conclusion, the states $\rho_{b,d}$ are entangled for $d\geq 4$
and $b\in(0,1)$.

Let us finally consider the missing cases of $d=2,3$ or $b=1$ or
$b=1$. For $d=2$ or $d=3$, theorem \ref{thm_PPT} says that
$\rho_{d,b}$ are PPT for any $b$. It is known
\cite{Horodecki96PLA} that all qubit-qubit and qubit-qutrit PPT
states are separable.

For $b=0$ it follows from Eq. (\ref{states}) that
$\rho_{d,0}=\proj{\Phi_0}$, which is separable, while $\rho_{d,1}$
can be written in the following separable form (cf. Ref.
\cite{Horodecki97PLA}):
\begin{equation}
\rho_{d,1}=\frac{1}{16\pi}\int_{0}^{2\pi}\mathrm{d}\varphi\,P(\varphi)\ot
Q(\varphi),
\end{equation}
where $P(\varphi)$ and $Q(\varphi)$ are projectors onto
$(1/\sqrt{2})(1,\mathrm{e}^{\mathrm{i}\varphi})$ and
$(1/\sqrt{d})(1,\mathrm{e}^{-\mathrm{i}\varphi},
\mathrm{e}^{-2\mathrm{i}\varphi},\ldots,\mathrm{e}^{-(d-1)\mathrm{i}\varphi})$,
respectively.
\end{proof}

\begin{thmm}
$r(\rho_{d,b})=r(\rho_{d,b}^{T_A})=d+1$.
\end{thmm}
\begin{proof}
Direct check shows that the vectors
\begin{eqnarray}
\ket{\Psi_i}&=&\ket{0,i}-\ket{1,i+1}\qquad
(i=1,\ldots,d-2)\nonumber\\
\ket{\Psi_i}&=&-\sqrt{1+b}\ket{00}+\sqrt{1-b}\ket{0,d-1}+\sqrt{1+b}\ket{11}.\nonumber\\
\end{eqnarray}
belong to the kernel of $\rho_{d,b}$ and the subspace they span
has dimension $d-1$. On the other hand, the product vector from
$R(\rho_{d,b})$, given in Eqs. (\ref{Prvecs1}) and (\ref{Prvecs2})
span $(d+1)$-dimensional subspace. Consequently,
$r(\rho_{d,b})=d+1$ and, since $\rho_{d,b}^{T_A}=(\mathbbm{1}_2\ot
U)\rho_{d,b}(\mathbbm{1}_2\ot U^{\dagger})$,
$r(\rho_{d,b}^{T_A})=d+1$.
\end{proof}


\begin{thebibliography}{0}



\bibitem{HReview}R. Horodecki {\it et al.}, Rev. Mod. Phys. {\bf 81}, 865 (2009).

\bibitem{teleportation}C. H. Bennett {\it et al.}, Phys. Rev. Lett. {\bf 70}, 1895 (1993).

\bibitem{Ekert}A. K. Ekert, Phys. Rev. Lett. {\bf 67}, 661 (1991).

\bibitem{Barrett}J. Barrett {\it et al.}, Phys. Rev. A {\bf 71}, 022101 (2005).

\bibitem{separable}R. F. Werner, Phys. Rev. A {\bf 40}, 4277 (1989).

\bibitem{Gurvits}L. Gurvits, {\it Classical deterministic complexity of Edmonds problem and quantum entanglement},
In Proc. of the 35th ACM symp. on Theory of comp., p. 10-19 (New
York, 2003, ACM Press).

\bibitem{SmolinExp}E. Amselem and M. Bourennane, Nat. Phys. {\bf 5}, 748 (2009); J. Lavoie {\it et al.}, Phys. Rev. Lett. {\bf 105}, 130501 (2010).

\bibitem{DickeExp}R. Prevedel {\it et al.}, Phys. Rev. Lett. {\bf 103}, 020503 (2009); W. Wieczorek {\it et al.}, Phys. Rev. Lett. {\bf 103}, 020504 (2009).

\bibitem{GHZExp}Y.-F. Huang {\it et al.}, Nat. Comm. {\bf 2}, 546 (2011).

\bibitem{DurCirac}W. D\"ur and J. I. Cirac, Phys. Rev. A {\bf 61},
042314 (2000).

\bibitem{multWit}A. Ac\'in {\it et al.}, Phys. Rev. Lett. {\bf 87}, 040401 (2001); G. T\'oth, Phys. Rev. A {\bf 71}, 010301(R) (2005).

\bibitem{DIW}J.-D. Bancal {\it et al.}, Phys. Rev. Lett. {\bf 106}, 250404 (2011).

\bibitem{SpinSq}J. K. Korbicz, J. I. Cirac, and M. Lewenstein, Phys. Rev. Lett. {\bf 95}, 120502 (2005);
G. T\'oth {\it et al.}, Phys. Rev. Lett. {\bf 99}, 250405 (2007).

\bibitem{SeevinckUffinkGuhne}M. Seevinck and J. Uffink, Phys. Rev. A {\bf 76}, 042105 (2007);
O. G\"uhne and M. Seevinck, New J. Phys. {\bf 12}, 053002 (2010).

\bibitem{GezaReview}O. G\"uhne and G. T\'oth, Phys. Rep. {\bf 474}, 1 (2009).

\bibitem{EckertAnP}K. Eckert {\it et al.}, Ann. Phys. {\bf 299}, 88 (2002).

\bibitem{Benatti}F. Benatti, R. Floreanini, and U. Marzolino, Ann. Phys. (NY)
\textbf{327}, 1304 (2012).


\bibitem{sym}A. R. Usha Devi, R. Prabhu, and A. K. Rajagopal, Phys. Rev. Lett. {\bf 98}, 060501 (2007).

\bibitem{Perm1}G. T\'oth and O. G\"uhne, Phys. Rev. Lett. {\bf 102}, 170503 (2009).

\bibitem{Perm2}G. T\'oth and O. G\"uhne, Appl. Phys. B {\bf 98}, 617 (2010).

\bibitem{symm}M. Hayashi {\it et al.}, Phys. Rev. A {\bf 77}, 012104 (2008);
T. Ichikawa, T. Sasaki, I. Tsutsui, and N. Yonezawa, Phys. Rev. A
\textbf{78}, 052105 (2008); T. Bastin {\it et al.}, Phys. Rev.
Lett. {\bf 103}, 070503 (2009); R. H\"ubener {\it et al.}, Phys.
Rev. A {\bf 80}, 032324 (2009); P. Mathonet {\it et al.}, Phys.
Rev. A {\bf 81}, 052315 (2010); C. D. Cenci {\it et al.}, Quant.
Inf. Comp. {\bf10}, 1029 (2010); D. J. H. Markham, Phys. Rev. A
{\bf 83}, 042332 (2011).

\bibitem{edge}M. Lewenstein {\it et al.}, Phys. Rev. A {\bf 63}, 044304 (2001).

\bibitem{Karnas}S. Karnas and M. Lewenstein, Phys. Rev. A {\bf 64}, 042313 (2001).

\bibitem{edgePapers}A. Sanpera, D. Bruss, and M. Lewenstein, Phys. Rev. A {\bf 63}, 050301 (2001);
S.-H. Kye and H. Osaka, J. Math. Phys. \textbf{53}, 052201 (2012)
and references therein.

\bibitem{Horodecki96PLA}M. Horodecki, P. Horodecki, and R. Horodecki, Phys. Lett. A {\bf223}, 1 (1996).

\bibitem{Jordi}R. Augusiak, J. Tura, J. Samsonowicz, and M.
Lewenstein, \textit{Entangled symmetric states of N qubits with
all positive partial transpositions}, arXiv:1206.3088.

\bibitem{2N}B. Kraus {\it et al.}, Phys. Rev. A {\bf 61}, 062302 (2000).

\bibitem{MN}P. Horodecki {\it et al.}, Phys. Rev. A {\bf 62}, 032310
(2000).



\bibitem{Normal}J. Samsonowicz, M. Ku\'s, and M. Lewenstein, Phys.
Rev. A {\bf 76}, 022314 (2007).



\bibitem{Horodecki97PLA}P. Horodecki, Phys. Lett. A {\bf 232}, 333 (1997).

\bibitem{Chruscinski}D. Chru\'sci\'nski and A. Rutkowski, Phys.
Lett. A {\bf 375}, 2793 (2011).
















\bibitem{Leinaas}J. M. Leinaas, J. Myrheim, and E. Ovrum, Phys. Rev. A {\bf 76}, 034304 (2007).

\bibitem{OptComm}R. Augusiak {\it et al.}, Opt. Comm. {\bf 283}, 805 (2010).









\end{thebibliography}
\end{document}